**(Preprint) IAA-AAS-SciTech2020-018**

# INTERPLANETARY CHALLENGES ENCOUNTERED BY THE CREW DURING THEIR INTERPLANETARY TRANSIT FROM EARTH TO MARS

**Malaya Kumar Biswal M[*] and Ramesh Naidu Annavarapu[†]**

Mars is the next destination after Earth to support terrestrial life. Decades of Mars exploration has fascinated space explorers to endeavour for a human expedition. But human Mars enterprise is complicated than conventional mission as the journey is endowed with a profusion of distinct challenges from terrestrial planet to the planetary surface. To perceive and overcome the implications of interplanetary challenges, we conducted a study to manifest every challenge encountered during interplanetary transit from Earth to Mars. Our study concluded entire challenges were attributed to the options for trajectory correction and maneuvering, management of space vehicles, the hazards of exposure to galactic radiation, effects of crew health in a microgravity environment, deficit solar power production, hazards of nuclear elements, psychologic and health effects, interrupted communication interlink from the ground, the complication in fuel pressurization and management, recycling of space wastes, execution of the extra-vehicular activity, and Mars orbital insertion. The main objective of this paper is to underline all possible challenges and its countermeasures for a sustainable crewed mission beyond low earth orbit in forthcoming decades.

## INTRODUCTION

Mars is the next frontier after Earth to support terrestrial life. The progression of rocketry and space science has exaggerated to send autonomous spacecrafts to explore Mars and beyond. In this current state of affair making humans, a multi-planetary species has become a goal and the next step of human civilization beyond low-earth orbit (Reference 1). Travelling to multiple destinations in our solar system affords a greater opportunity to determine the extent of human presence in space, and to demonstrate the extent of scientific technology. Further it enables us to understand the origin and evolution of life in our vast galaxy or solar system. However, Mars excursion is not an easy task and are endowed with numerous diverse challenges beginning from our terrestrial planet up to the destination point. Concerning sustainable and a prosperous human class mission, we have provided and underlined every possible challenge and their implications encountered to the crew during their interplanetary transit from the Earth to Mars. Outline Map for overall interplanetary challenges is shown in Figure 1.

---

[*] Graduate Researcher, Department of Physics, Pondicherry University, Kalapet, Puducherry – 605 014, India. Student Member of AIAA and Indian Science Congress Association, India. Email: **malaykumar1997@gmail.com; mkumar97.res@pondiuni.edu.in**.
[†] Associate Professor, Department of Physics, Pondicherry University, Kalapet, Puducherry – 605 014, India.
Email: **rameshnaidu.phy@pondiuni.edu.in; arameshnaidu@gmail.com**.

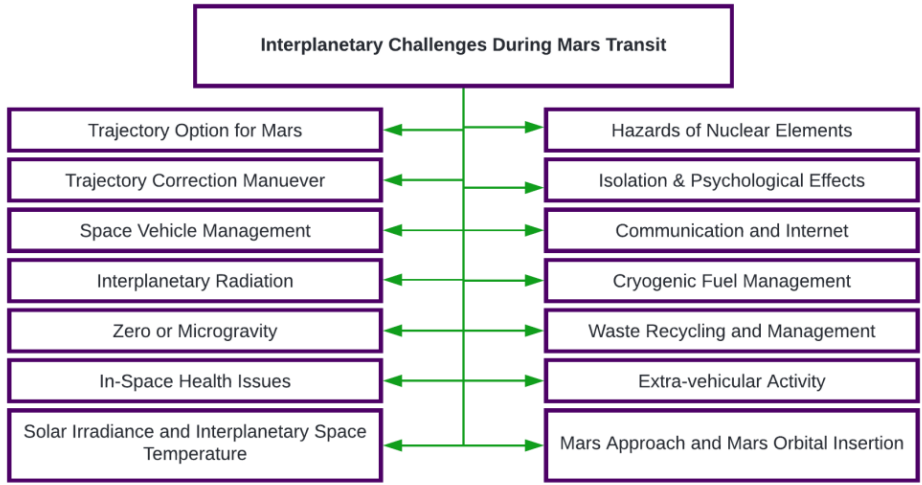

**Figure 1 Outline Map for Interplanetary Challenges**

## TRAJECTORY OPTION FOR MARS TRANSIT

Once we left out from low-earth orbit, we will be having two trajectory choices: either we have to pick conjunction class or opposition class (References 2-5). But trajectory analysts have optimized and proposed numerous trajectory architectures for the interplanetary cruise to Mars. Here we have considered two major class trajectories for discussion.

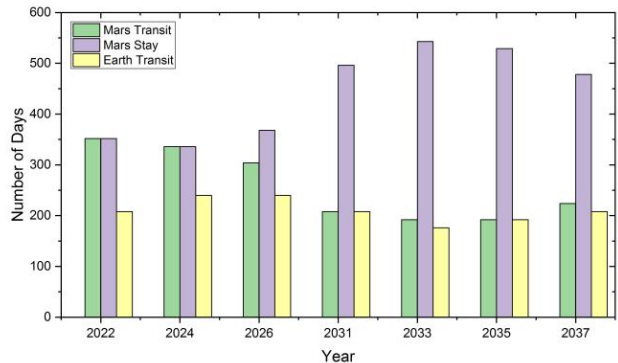

**Figure 2 Duration of Missions for Distinct Launch Window**

### Conjunction Class Trajectory

Conjunction class trajectory is often referred to as long-stay approach or is best known for longer stay missions. Since astronauts have most of their mission time on Mars (i.e. 400-600 days). Therefore, longer stay mission ensures crew safety than opposite class trajectory because longer stay allows the crew to get sheltered under the Martian environment than from being exposed to harmful cosmic radiation. In addition to this, mission planners get benefited by this type of trajectory due to the planetary alignment of the Earth and Mars. Hence crewed space vehicle grasps a minimal delta velocity of approximately 3.36 km/s to follow Hohmann's transfer trajectory to approach Mars. Thus the conjunction class trajectory affords a simplest and safest pathway to approach Mars. Further, conjunction class trajectory enables easiest Mars Orbit Insertion due to its optimal delta velocity (Reference 6).

**Opposition Class Trajectory**

Contradictory to conjunction class, opposition class trajectory is known as a short-stay mission where astronauts have most of their mission time in interplanetary transit than the surface of Mars. This class of trajectory allow crews to stay on the planet with a minimum duration of 30 days to a maximum extent of 60 days. Opposition class trajectory is found to be the riskiest approach than conjunction class because it consumes more energy to give out maximum delta velocity of 7 km/s and longer duration in interplanetary space is subjected to the vulnerability of galactic and intergalactic cosmic radiation. In addition to this, the delta velocity of 7 km/s requires maximized backward propulsion (that consumes maximum fuel) to enable safer Mars Orbital Insertion. Further, the requirement or necessity of Venus flyby may cause increased vulnerability to Sun's hazardous elements due to their proximity passage (Reference 6).

**Trajectory Assessment for Mars Transit**

Several cases of studies show that opposition class trajectory has more vulnerability to the hazardous cosmic radiation and may cause serious health effects. So staying longer duration on Mars under natural radiation shield (Mars atmosphere) is much safer than spending time in interplanetary space. Effectively the hazard of radiation exposure can be minimized on Mars by the application of Mars Sub-Surface Habitats or using Martian regolith as roof layer of deep space habitats. In addition to this, preferring conjunction class and its longer surface stay can be productively spent in gathering results of more scientific interest from various scientific sites. Since human mission to Mars cost expensive than conventional missions and limits the availability of regular cargo transit from Earth and Mars or vice versa. Similarly, the opposition class trajectory may limit the crew to execute long term experiments. Henceforth, from overall analysis and the perspective of crew safety, we recommend the conjunction class trajectory which is ideal for the human-crewed mission as it affords crew safety, cheaper mission, and effective of all aspects. Optimized trajectory for Mars launch window of between 2020 and 2040 is shown in Figure 3.

**Mission Abort and Back-up Plan**

In case of any critical or an emergency situation, the crew can abort the mission, vast-off Mars and can return to Earth using either high energy optimal trajectory path or Mars free-return trajectories (References 7 and 8). One of the simplest optimized trajectory simulated from Trajectory browser tool between the time frame of 2020 and 2040 is shown in Figure 1. And the duration required for both Mars and Earth transit and Mars stay is shown in Figure 2 (Reference 9).

**TRAJECTORY CORRECTION MANEUVERING**

Failure analysis on conventional Mars probes shows that one-fourth of the spacecraft encounters constraints in igniting the manoeuvre correction engines that caused as a result of damage of the thermal control system of the spacecraft (References 10 and 11). So a voyage to Mars along the interplanetary coast is subjected to low-temperature and pressure, and zero-gravity environment. And these conditions directly create an impact on the maintenance of fuel and retaining zero-boil-off storage of cryogenic fuel. Therefore, inadequate fuel management may give rise to the inappropriate firing of course-correction thrusters allocated for trajectory correction maneuver. In addition to this, greater delta velocity either during the Earth or Mars departure may have serious concerns over maneuvering massive space vehicle. Hence, adequate fuel pressurization and management with optimal delta velocity may cut-off these challenges (Reference 12).

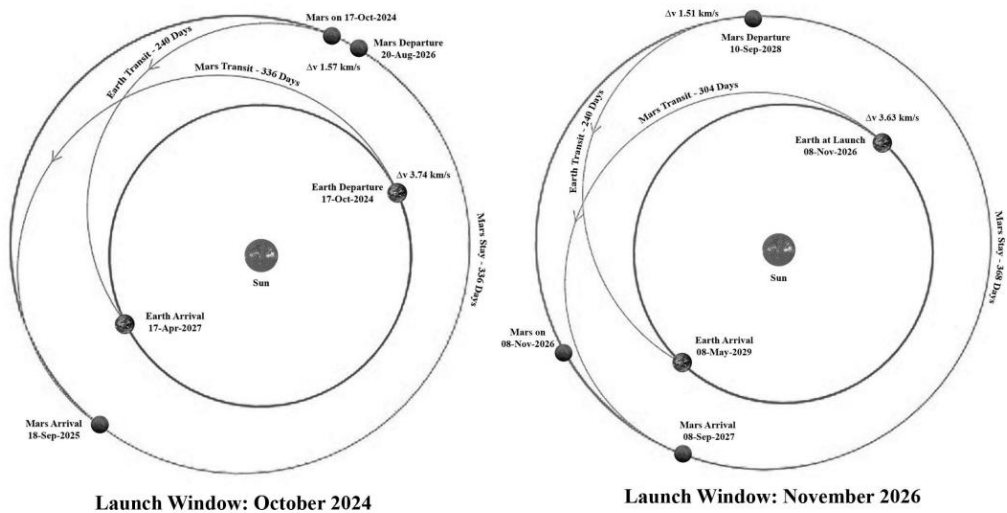

**Figure 3 Optimized Trajectory for Human Mission to Mars**[*]

**MANAGEMENT OF SPACE VEHICLE**

Space vehicle management and maintenance are some of the basic and challenging tasks for the travelling astronauts. It is because sustaining under a zero-gravity or microgravity environment may give rise to several physical health issues and limiting their access over complete space vehicle. As we know that the electronic components are degraded when exposed to cosmic radiation (Reference 13) and the internal damages to the electronic components of the space vehicle can be minimized with servicing crews, but external damages are quite challenging. Hence the construction of space vehicles with durable and robust electronics are decidedly recommended. Further, it is desirable to employ automated robots for detecting damage and space vehicle management (Reference 14).

**EFFECT OF RADIATION AND ZERO-GRAVITY**

**Radiation**

Radiation and Zero-gravity are the confronting challenges of human spaceflight or interplanetary spaceflight. On an interplanetary transit to Mars, astronauts are greatly exposed to the galactic and intergalactic cosmic radiation, solar cosmic rays, and solar particles events. These are the natural phenomena spontaneously originating from our galaxy and pose a threat to spaceflight safety systems. National Aeronautics and Space Administration (NASA) has identified major health issues of GCR and SCR exposure. They can be categorized as carcinogenesis, cardiovascular disease, tissue degeneration, damage to the central nervous system, and acute radiation syndrome. Hence concerning the crew health, it is very significant to have good radiation suit aboard space vehicle for a durable mission (References 15 and 16).

---

[*] NASA Ames Research Center Trajectory Browser. Trajectory between Earth and Mars. Available Online at https://trajbrowser.arc.nasa.gov/traj_browser.php accessed 06 September 2020.

In addition to health issues, space radiation can cause unrecoverable damage to the electronic components or onboard circuitry systems of the space vehicle that may result in spacecraft malfunction. So, the adequate thickness of spacecraft radiation suit is considered to avoid radiation damages. Because interplanetary transit to Mars may lead to longer duration exposure to space radiation for about 180-270 days. The radiation dosage has an average of 1.16 millisieverts in interplanetary space, similarly, the dosage is high in case of opposition class trajectory due to the additional Venus flyby (References 16 and 17). Dosage of Radiation exposure in interplanetary space is shown in Figure 4.

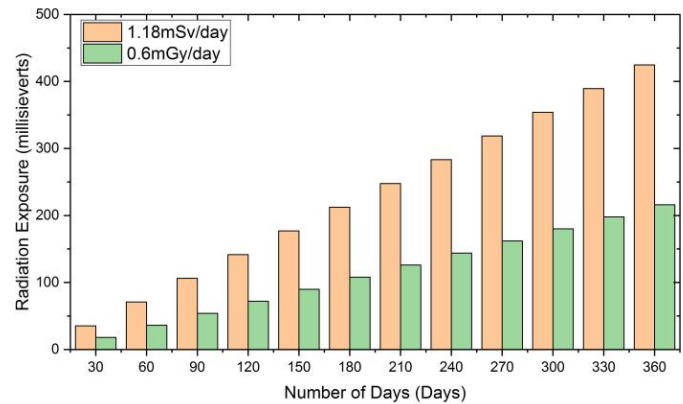

**Figure 4 Radiation Exposure in Interplanetary Space**

**Zero or Micro-gravity**

Astronauts during interplanetary spaceflight are subjected to three sorts of gravitational field; first when they are transiting between the planets, second is during the surface stay on Mars and the third is while returning back to Earth. These three sorts of gravitation transition may cause improper coordination of brain functions, improper spatial balance, and motion sickness. NASA has been performing several experiments aboard International Space Station (ISS) to understand the effect of microgravity on human health. Experiments showed that the crew suffers from the collapse of bone density due to the loss of bone minerals (osteoporosis), inadequate intake of necessary consumables and lack of regular exercise have caused the loss of muscle endurance[*] (Reference 18).

**Historical Observation of the effect of Zero-gravity on human health**

Observation and study from past Space Shuttle Program, Mir, and ISS Expedition showed that the astronauts had severe health impact on their bones, muscle systems, and cardiovascular system (Reference 19). Similarly, upon returning from the Space Station the crew had the additional implication of improper blood pressure and blood circulation to the brain. Hence human mission beyond low-earth orbit may be subjected to long exposure to a microgravity environment and posing a challenge to space rehabilitation. So, these challenges can be addressed by generating artificial gravity aboard space vehicle and practicing adequate food habitation along with good physical exercise.

---

[*] Abadie, L (2015). Gravity, who needs it? NASA studies your body in space. Accessed from https://www.sciencedaily.com/releases/2015/11/151118155434.htm on 06 September 2020.

**ISSUES OF SOLAR IRRADIANCE AND TEMPERATURE**

Solar irradiance has a significant role in the generation of solar power required for powering the spacecraft and regulation of stabilized temperature to maintain thermal stability. Deficiency of electrical power due to lower solar irradiance may reduce the efficiency of the operating devices aboard space module. Since the availability and intensity of solar irradiance gradually decrease as we move far from the center of the solar system. The solar irradiance varies in the order of 1366W/m2 at Earth to 588 W/m2 at Mars (The record is estimated as per solar energy generation capacity of NASA's Mars Reconnaissance Orbiter) (Reference 20). Distribution of Intensity of Solar Irradiance over solar system is shown in Figure 5.

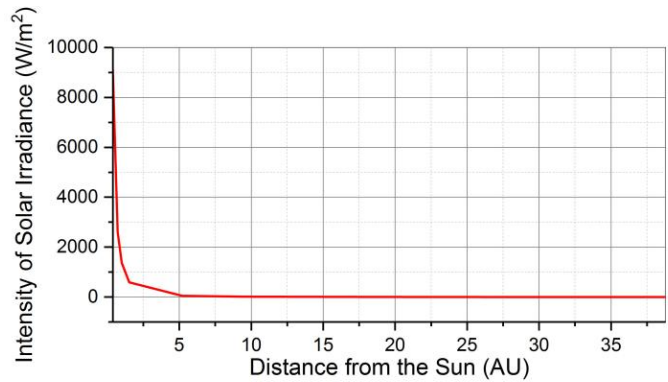

**Figure 5 Distribution of Solar Irradiance in the Solar System**[*]

Parallel to the solar irradiance the temperature of the interplanetary space falls as a function of inverse square law. The temperature variance affects the thermal control system and zero boil-off storage of spacecraft propellant. Further, the temperature imbalance may result in damage of electronic components due to thermal radiation and damage of life support systems aboard space vehicle. Temperature Fall of the Interplanetary Space is shown in Figure 6.

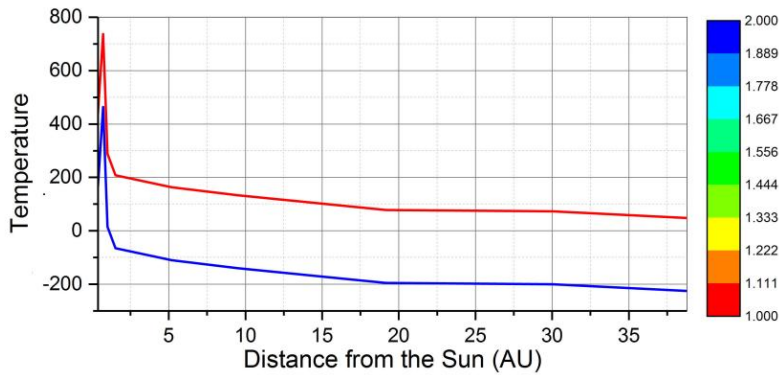

**Figure 6 Distribution of Temperature of Interplanetary Medium (Blue line in Celsius scale and Redline in Kelvin scale)**[*]

---

[*] Christiana Honsberg and Stuart Bowden. Solar Radiation in Space. Accessed from https://www.pveducation.org/pvcdrom/properties-of-sunlight/solar-radiation-in-space on 06 September 2020.

**EFFECT OF NUCLEAR HAZARDS ON CREW HEALTH**

Nuclear energy is an emerging technology for the future propulsion system. Several mission designers have proposed to employ nuclear thermal propulsion or nuclear electric propulsion to minimize the duration of interplanetary spaceflight for a faster mission. And nuclear power is an ideal choice that can meet the strategy of space exploration beyond LEO. But the hazardous nuclear elements from nuclear reactors of space vehicle can cause severe health effects that we have discussed in the section "**Radiation**". So, we recommend mounting the crewed module at a far and safer distance from the nuclear reactor / nuclear propulsion systems (References 21 and 22)

**ISOLATION AND PSYCHOLOGICAL EFFECTS ON CREW**

A mission to Mars takes an average of 2-3 years and that may push the astronauts into a complete state of isolation. The long state of confinement affects the psychological and behavioural patterns. As part of behavioural changes, the crew may encounter issues concerning mood and cognition, the risk of anxiety and depression, digestive problems, hypotension or hypertension (Reference 23), and loneliness. Similarly, as part of psychological changes astronaut may suffer from improper positive moods and relationship pattern with onboard crew members. Some analog simulation experiments on human habitation (Mars 500) shows an increase in positive moods among crews (Reference 24), but this results may remain undesirable as the actual space environment is dissimilar and cannot be compared with the simulated environment on the Earth. But somehow, the results and experiences from analog stations may provide a useful framework for future missions.

In addition to this, the experiences and lessons learned from International Space Station can be considered as a technical guide or vision for future interplanetary transit. Further, the astronauts may encounter communication relay from the ground, family, and friends that create a direct impact on crew personality behaviour. It is because astronauts have to wait for 30-40 minutes for both transmitting and receiving a single message and that can lead to critical stressful situations (Reference 25). These challenges are cannot be avoided and are far beyond how well trained or experienced the crew are. Therefore, it is recommended to select astronauts who are physically and mentally fit in terms of psychology, and they must be enriched with multidisciplinary skills. So that the crew can manage themselves during stressful situations and can perform multiple tasks at any emergencies (Reference 26).

**DELAYED COMMUNICATION AND INTERPLANETARY INTERNET**

**Delayed Communication**

In-space communication plays a major role in engaging the mission and the mission crew and also keeps the astronauts updated about the mission plans. Parallel to this, it enhances the psychological personality of the crew. So advanced communication is highly essential because at a distance of 1.5-2.5 AU the communication interlink is limited to 24-40 minutes. And it may not be an effective approach to stay interlinked with the crew and operating Mars relay orbiters (References 27 and 28).

The human mission to Mars is completely new and the astronauts are exposed to a new inexperienced environment. So it is substantial to keep the astronauts updated about the mission strategies, safety measure and their next move. Furthermore, unavailability of continuous communication interlinks over the manned vehicle during the blackout period of solar conjunction, the technological necessity to have control over multiple proximity operations and the demand for wider coverage of deep space network are some of the challenges on the way to communication relay systems (Reference 29).

---

[*] Williams, D.R (2016). NASA Space Science Data Coordinated Archive. Planetary Fact Sheets. National Aeronautics Space Administration. Accessed from https://nssdc.gsfc.nasa.gov/planetary/factsheet/ on 06 September 2020.

**Interplanetary Internet**

Interplanetary Internet is a concept of providing internet access to the travelling astronauts. This may enable the crew to access and perform a wide range of interdisciplinary studies with massive access to reliable scientific resources. In addition to this, it may allow the public to get updated or to get live coverage from interplanetary space. This mode of tethering directly from the public may reduce the crew mental stress and increase psychological health. Modern networking technology can able to achieve a maximum downlink of ~100-2015 Mbps and an uplink of ~10-25 Mbps via Earth-Mars trunk line (Reference 29). Among both interlinks, the enhancement of uplink transfer is highly required. Currently, NASA is planning to improve the communication architecture by parking two-three communication relay satellites in HMO or to park satellites in Earth-Sun Lagrangian point with the additional capability of optical fibre and laser-guided communication system (Reference 30).

## SIGNIFICANCE OF CRYOGENIC FUEL MANAGEMENT

In the case of the chemical propulsion system for maneuvering space vehicle, cryogenic fuel management is necessary for the extended presence in space. It is because the propellant is the driving force of the spacecraft and helps in transiting from the Earth to Mars as well as to perform course corrections. Improper fuel management may result in fuel depletion and loss of mission. Further, the future manned exploration greatly relies on three modes of the extraterrestrial cryogenic fuel management system, they are the propellant management, storage, and distribution. Several other factors that affect CFM that is clearly described in (References 31 and 32). To enhance and demonstrate the extent of cryogenic fuel management technology, NASA has performed various experiments in space (i.e. CPOD, MDSCR, ZBOT, ISCPD) (References 32 and 33) and are found to be effective for CFM in future missions.

## SPACE WASTE RECYCLING AND MANAGEMENT

Longer presence of crews over interplanetary spaceflight may lead to exhaustion of resources and generation of space wastes. The space waste are of multiple categories posing a threat to the crew as a biological and physical hazard. The type of trashes generated includes $CO_2$ gases (exhaled by the astronauts), used waters, human wastes, solid wastes, nuclear and medical wastes. Among these, biological, medical, and nuclear waste has the potential to put the crew to prolonged cancer sickness. So, disposal or management of this kind of waste includes confined and proper disposal procedure with an anaerobic digester (Reference 34). An estimate shows that a crew of four can generate trash up to 2.5 tons per year, and hence the crew over 3 years may able to generate 7-8 tons (Reference 35). So, we recommend that implementing recycling methods rather than conventional disposal procedure is worth effective to avoid the concerns of cargo re-supply. Further recycling is an ideal approach to compensate for the limit of cargo re-supply from the terrestrial planet. The current state of recycling procedures aboard ISS can be improved and applied to longer transit duration to Mars (References 36 and 37) and future deep space transportation systems.

**EXTRA-VEHICULAR ACTIVITY AND SPACE WALKS**

Extra-vehicular activity is highly essential to promote space activities of astronauts outside their confined environment. It determines the human's potential capability for space exploration and the duration of extra-vehicular performance. But the challenge like retaining constant pressure and temperature within the suit, oxygen concentration, capability of supplying food and water, and waste collection pose a technical challenge to EVA. In addition to this, the threat of exposure to cosmic radiation, the challenge in misbalancing crew's momentum in microgravity, mechanical hazard, and the risk of losing in space are the confronting challenges of EVA (Reference 38).

Hence, we recommend that astronaut with good EMU (Extra-vehicular maneuvering Unit) suit may decrease the risk of losing in space. Moreover, the fabrication of durable and sufficient radiation-proof suit may protect the astronauts from radiation and mechanical hazard. Currently, NASA's Johnson Space Center is developing Robonaut to assist and enhance EVA with the improved capability of strength and mobility (References 39-42).

**MARS APPROACH AND MARS ORBITAL INSERTION**

The destination of interplanetary transit is Mars. So after a transit duration of 6-9 months in interplanetary space, it is very significant to achieve successful Mars Orbital Insertion[*]. Because once if we miss the targeting planet it will be a more complementary task to maneuver back the space vehicle to its destination. Additionally, the factors such as optimal delta velocity, adequate fuel management, proper functioning of navigation and maneuvering system determine the success rate of Mars Orbital Insertion (MOI). Further upon successful Mars orbit capture the crew may encounter the hazard of harmful cosmic radiation and threat of asteroid impact. So, the cosmic hazard can be minimized by directing the crew module to land on Mars thereby leaving the space vehicle in Mars orbit. The atmosphere of Mars ensures crew safety from space radiation and the threat of asteroid impact. Radiation levels of both Earth and Mars orbit is shown in Figure 7.

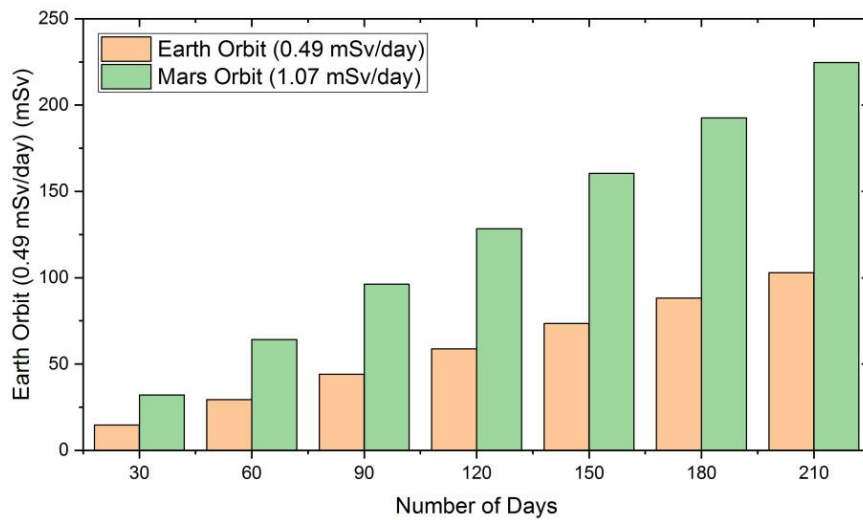

**Figure 7 Comparison of Radiation Exposure in Earth and Mars Orbit[*]**

---

[*] NASA. What is Mars Orbit Insertion? Accessed from https://www/mars.nasa.gov. on 06 September 2020.

## CONCLUSION

Considering overall challenges on the way to the human expedition to Mars, we have selectively reviewed the challenges encountered during interplanetary transit from Earth to Mars. In addition to diverse challenges, we have discussed and recommended some approaches to redress the challenges in appropriate section. Overall challenges of human mars exploration and expedition are technically and briefly reviewed in (References 43 and 44) Our study is the outcome of past experiences and lesson learned from the past forty-four conventional Mars probes. Finally, we expect that our study may provide some useful framework for the explorers and future space travelers to overcome the interplanetary challenges.

## ACKNOWLEDGEMENTS

I (Malaya Kumar Biswal) would like to thank my supervisor, **Prof. A. Ramesh Naidu. Ph.D.** (author), for the patient guidance, encouragement and advice he has provided throughout my time as his student. Further I would like to extend my sincere thankfulness to all of my lovable friends for their financial support for the conference participation.

## CONFLICT OF INTEREST

The authors have no conflict of interest to report.

## DEDICATION

I (Malaya Kumar Biswal) would like to dedicate this work to my beloved mother late **Mrs. Malathi Biswal**, father **Mr. Madhav Biswal**, and my family.

## ACRONYMS

| | | |
|---|---|---|
| AU | - | Astronomical Unit |
| CFM | - | Cryogenic Fuel Management |
| CPOD | - | Cryogenic Propellant Operations Demonstrator |
| EMU | - | Extra-Vehicular Maneuvering Unit |
| EVA | - | Extra-Vehicular Activity |
| GCR | - | Galactic Cosmic Radiation |
| HMO | - | High Mars Orbit |
| ISS | - | International Space Station |
| LEO | - | Low Earth Orbit |
| MBPS | - | Mega Bytes Per Second |
| MDSCR | - | Maturation of Deep Space Cryogenic Refueling Technology |
| MOI | - | Mars Orbital Insertion |
| MRO | - | Mars Reconnaissance Orbiter |
| NASA | - | National Aeronautics and Space Administration |
| ISCPD | - | In-Space Cryogenic Propellant Depot |
| SCR | - | Solar Cosmic Rays |
| ZBOT | - | Zero Boil-Off Tank |